\documentclass[twocolumn,superscriptaddress,prl,aps,floats,showpacs]{revtex4}
\usepackage{amsmath,amssymb,amsfonts,amsthm}
\usepackage{graphicx}
\usepackage{natbib}
\usepackage{bm}
\newcommand{\Var}{\mathrm{Var}}
\newcommand{\Cov}{\mathrm{Cov}}

\begin{document}
\title{Duality and KPZ in Liouville Quantum Gravity}
\author{Bertrand Duplantier}
\affiliation{Institut de Physique Th\'{e}orique, CEA/Saclay, F-91191
Gif-sur-Yvette Cedex, France}
\author{Scott Sheffield}
\affiliation{Department of Mathematics,  Massachusetts Institute of Technology,
Cambridge, MA 02139}\altaffiliation{On leave of absence from the Courant Institute for Mathematical Sciences at NYU.}

\date{January 2, 2009}

\begin{abstract}
We present a (mathematically rigorous) probabilistic and geometrical
proof of the KPZ relation between scaling exponents in a Euclidean
planar domain $D$ and in Liouville quantum gravity. It uses the
properly regularized quantum area measure
$d\mu_\gamma=\varepsilon^{\gamma^2/2} e^{\gamma
h_\varepsilon(z)}dz,$ where $dz$ is Lebesgue measure on $D$,
$\gamma$ is a real parameter, $0\leq \gamma <2$, and
$h_\varepsilon(z)$ denotes the mean value on the circle of radius
$\varepsilon$ centered at $z$ of an instance $h$ of the Gaussian
free field on $D$. The proof extends to the boundary geometry. The
singular  case $\gamma > 2$ is shown to be related to the quantum
measure $d\mu_{\gamma'}$, $\gamma'<2$, by the fundamental duality
$\gamma\gamma'=4$.
\end{abstract}
\maketitle

\textit {Introduction.} One of the major theoretical
advances in physics over the past thirty years has been the
realization in gauge theory or string theory that transition
amplitudes
 require summing over random surfaces, which replaces traditional sums
over random paths, i.e., the celebrated Feynman path integrals of quantum mechanics and quantum field theory. In a 1981
breakthrough,
A.M. Polyakov \cite{MR623209,MR1122810} first understood that the summation
over random Riemannian metrics involved
 could be represented mathematically by the so-called \textit{Liouville theory of quantum gravity}.

 The latter can be simply described as follows:
Consider a bounded planar domain $D\subset \mathbb C$ as the parameter domain of the random
Riemannian surface, and an instance $h$ of the
Gaussian free field (GFF)
on $D$, with Dirichlet energy
$(h,h)_{\nabla}:=({2\pi})^{-1}
\int_D \nabla h(z) \cdot \nabla h(z)dz.$ The quantum area is then (formally) defined by
$\mathcal A=\int_D e^{\gamma h(z)}dz$, where
$dz$ is the standard Euclidean  (i.e., Lebesgue) measure and $e^{\gamma h(z)}$ the (random)
conformal factor of the Riemannian metric, with a constant $0\leq
\gamma <2$. The quantum Liouville action is then
\begin{equation}
S(h)=\frac{1}{2}(h,h)_{\nabla}+\lambda\mathcal A,
\label{liouville}
\end{equation}
where $\lambda \geq 0$ is the so-called ``cosmological constant''.

In 1986, V.A. Kazakov introduced the idea of placing a (critical) statistical model on a random planar lattice, when exactly solving there
the Ising model \cite{1986PhLA..119..140K}.
This led to the 1988 breakthrough by Knizhnik, Polyakov and Zamolodchikov \cite{MR947880}, building on
\cite{Polyakov:1987zb}, who predicted that
corresponding
critical exponents (i.e., conformal weights $x$) in the Euclidean plane and in quantum gravity ($\Delta$) would obey the
KPZ relation:
\begin{equation}
x=\frac{\gamma^2}{4}\Delta^2+\left(1-\frac{\gamma^2}{4}\right)\Delta.
\label{KPZrelation}
\end{equation}
In the critical continuum limit, the statistical system born by the random lattice is
described by a conformal field theory (CFT) with central charge $c\leq 1$, which fixes
$\gamma=\left(\sqrt{25-c}-\sqrt{1-c}\right)/\sqrt{6} \leq 2$ \cite{MR947880}. In Liouville quantum gravity
the CFT action is then simply added to
the Liouville one (\ref{liouville}), allowing an alternative derivation of (\ref{KPZrelation}) in the so-called conformal gauge \cite{1988MPLA....3.1651D,MR1005268}.

This provides the core continuous model of ``two-dimensional quantum gravity'', whose connections to string theory and
conformal
field theory, and to random planar lattice models and random matrix theory are manifold; see, e.g., the surveys
\cite{MR1122810,1990PThPS.102..319S,Ginsparg-Moore,MR1461284,MR1465433,2004IJMPA..19.2771N,MR2112128}
and references therein.

Despite the fact that the original work by KPZ has been cited almost a thousand times, and relation (\ref{KPZrelation}) repeatedly
checked, in particular by explicit calculations in geometrical models on random planar lattices
\cite{1988PhRvL..61.1433D,1989MPLA....4..217K,1992NuPhB.386..520K,MR1666816}, the KPZ relation was not proven rigorously until very recently.
In the probabilistic and geometrical approach of \cite{2008arXiv0808.1560D}, it is proven by starting
from the  \textit{critical}
 Liouville gravity, with action
$S$
(\ref{liouville}) taken at $\lambda=0$, i.e., a \textit{free-field} action,
and (\ref{KPZrelation}) appears as a consequence
of the fine properties of the two-dimensional GFF. Several follow-up works appeared since,
either at the rigorous level \cite{2008arXiv0806.1347B,2008arXiv0807.1036R} or heuristic one \cite{2008arXiv0810.2858D}.

One striking and important consequence of
our perspective is that KPZ holds not only within the original CFT realm, but for \textit{any
fractal structure as measured with the random metric or stochastic measure $e^{\gamma h(z)}dz$, and  for any $0\leq
\gamma <2$}. For instance, it suggests that the same Euclidean exponent $x$ of a random or self-avoiding walk
  would obey (\ref{KPZrelation})
with $\gamma=\sqrt{8/3}$ in pure gravity ($c=0$), but with $\gamma=\sqrt{3}$ on a random lattice equilibrated with Ising
spins ($c=1/2$), yielding different $\Delta$'s.

The aim of this Letter is to present this derivation in a minimal, yet rigorous way to the physics community.
It involves
using a properly defined and regularized quantum area measure, which allows for a transparent geometrical understanding
of KPZ. It utilizes only the underlying Brownian stochastic
 properties of the
 GFF, whose correlation functions can be expressed  in terms of
 the classical two-dimensional Newtonian potential.
We also prove the
boundary analog of KPZ for fractal subsets of the boundary $\partial D$ \cite{2008arXiv0808.1560D}.

Our probabilistic approach also allows us to consider
the \textit{duality} property of Liouville quantum gravity:  for $\gamma > 2$, the \textit{singular} quantum
measure can be properly defined in terms of the \textit{regular} $\gamma'$-quantum measure, for the dual value
$\gamma'=4/\gamma <2$,
confirming  that the so-called
``other branch'' of the $\gamma$-KPZ relation corresponds to standard $\gamma'$-KPZ for $\gamma'<2$, as advocated long ago
by Klebanov \cite{1995PhRvD..51.1836K,1995NuPhB.434..264K,1996NuPhS..45..135K}.

\textit{GFF circular average and Brownian motion.} Let $h$ be a centered Gaussian free field on a
bounded simply connected domain $D$ with Dirichlet zero boundary conditions. As already
remarked in \cite{Polyakov:1987zb}, special care is required
to make sense of the quantum gravity measure, since the GFF is a distribution
and not a function (it typically oscillates between $\pm \infty$) (see e.g., \cite{MR2322706}).

\begin{figure}
\begin{center}
\includegraphics[angle=0,width=.4690\linewidth]{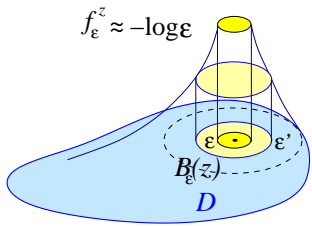}
\end{center}
\caption{\label{f.QG0} Potential $f_\varepsilon^z$ (\ref{Newtonian}) created by a uniform mass distribution on the circle
$\partial B_\varepsilon (z)$; its
leading constant value inside the disc $B_\varepsilon (z)$ is $-\log \varepsilon$.}
\end{figure}

For each $z \in D$, write $B_\varepsilon(z) = \{w: |w-z| < \varepsilon \}$.
When $B_\varepsilon(z) \subset D$, write $h_\varepsilon(z)$ for the average value of $h$ on the
circle $\partial B_\varepsilon(z)$.
Denote by $\rho_{\varepsilon}^z(y)$
 the uniform Dirac density (of total mass one) on the circle $\partial
B_\varepsilon(z)$, such that one can write the scalar product on $D$: $h_\varepsilon(z)=(h,\rho_{\varepsilon}^z):=
\int_D h(y)\rho_{\varepsilon}^z(y) dy$. To the density $\rho_{\varepsilon}^z$ is naturally associated a Newtonian
potential.
We define the function $f_\varepsilon^z(y)$, for $y \in D$,
\begin{equation}
 f_\varepsilon^z(y)=-\log \text{max}(\varepsilon, |z-y|)-\tilde G_z(y)
 \label{Newtonian}
\end{equation}
where $\tilde G_z(y)$ is the
\textit{harmonic} function of $y\in D$, with boundary value equal to the restriction of
$-\log |z-y|$ to $\partial D$. By construction
this $f_\varepsilon^z(y)$ satisfies Dirichlet b.c.'s
and the Poisson equation $-\Delta f_\varepsilon^z=2\pi \rho_\varepsilon^z$. This (regular) potential function is
represented in Fig.\ref{f.QG0}.
Integrating by parts, we immediately have the following:
\begin{equation}
h_\varepsilon(z) =(h,f_{\varepsilon}^z)_\nabla ,
 \label{hf}
\end{equation}
in terms of the Dirichlet inner product
$$(f_1, f_2)_\nabla := (2\pi)^{-1} \int_D \nabla f_1(z) \cdot \nabla f_2(z) dz,$$
i.e.,  the interaction energy of fields associated with potentials $f_{i=1,2}$.
In fact, the random variables  $(h, f)_\nabla$
 are zero mean
Gaussian random variables for each $f$, with the covariance property:
$\Cov\bigl((h, f_1)_\nabla,(h, f_2)_\nabla \bigr) = (f_1, f_2)_\nabla.$
From this, we deduce the covariance of the averaged $h_\varepsilon(z)$ fields (\ref{hf}) on two nested circles
(Fig.\ref{f.QG0}):
 $\Cov \bigl(h_\varepsilon(z),h_{\varepsilon'}(z)\bigr)=(f_{\varepsilon}^z,f_{\varepsilon'}^z)_\nabla.$
 The latter is the Newtonian interaction energy of the two circles, which,
 owing to the explicit potential (\ref{Newtonian}), to Gauss' theorem
and to harmonicity of $\tilde G_z$, gives the explicit form of the covariance
\begin{eqnarray}\mathbb E \bigl(h_\varepsilon(z)h_{\varepsilon'}(z)\bigr)
=  -\log \max (\varepsilon,\varepsilon') -\tilde G_z(z),
\label{hh'explicit}
\end{eqnarray}
with $\mathbb E\, h_\varepsilon(z) = 0$ for Dirichlet boundary conditions, and where $\tilde G_z(z)=-\log C(z; D)$
in terms of the so-called conformal radius $C$ of $D$
viewed from $z$, a smooth function of $z$.
From (\ref{hh'explicit}) we thus get the two important variances
\begin{eqnarray}\label{varh}
\Var\, h_\varepsilon(z)&=&-\log \varepsilon+\log C(z; D)\\
\Var \bigl[h_\varepsilon(z)-h_{\varepsilon'}(z)\bigr]
&=&  |\log \varepsilon-\log \varepsilon'|.
\label{varh-h'}
\end{eqnarray}
The interpretation of (\ref{varh},\ref{varh-h'}) is immediate: for fixed $z$, the Gaussian random variable
$h_\varepsilon(z)$ is  \textit{one-dimensional standard Brownian motion when parameterized by time
$t:=-\log\varepsilon$} \cite{2008arXiv0808.1560D}.

\textit{Random metrics and Liouville quantum gravity.}  Recall first that if $N$ is a
Gaussian random variable with mean $a$ and variance $b$ then
 $\mathbb E \, e^N =
e^{a+b/2}$. Since $\mathbb E\, h_\varepsilon(z) = 0$, we have from (\ref{varh})
the exponential expectation:
\begin{equation} \label{Eexphepsilon} \mathbb E\, e^{\gamma h_\varepsilon(z)}=e^{{\Var}[ \gamma
h_\varepsilon(z)]/2}
= \left[ {C(z;D)}/{\varepsilon} \right]^{\gamma^2/2}.
\end{equation} Since  (\ref{Eexphepsilon}) ultimately \textit{diverges} for $\varepsilon \to 0$,
we are led to \textit{regularize}  Liouville quantum gravity
by defining the random measure
\begin{equation}
\label{liouvillemeasure}
d\mu_\gamma=M_\varepsilon(z)dz, \,\,\,\,\,\,\, M_\varepsilon(z) := \varepsilon^{\gamma^2/2} e^{\gamma h_\varepsilon(z)},
\end{equation}
in a way similar to the so-called Wick normal ordering
 (see e.g., \cite{BSimon}). In \cite{2008arXiv0808.1560D}, it is shown that the limit of this regularized measure
 exists as $\varepsilon \to 0$, which mathematically defines  {Liouville quantum gravity} (see also \cite{MR0292433}).

\textit{GFF sampling and random metrics.}
We now consider a measure on pairs $(z,h)$ where $h$ is the Gaussian free field,
and given $h$ the point $z$ is chosen from the regularized quantum area measure
$e^{\gamma h_\varepsilon(z)}dz$.  Such a measure has the form $e^{\gamma h_\varepsilon(z)} dh dz$ where
$dh$ represents the (whole) GFF measure.
Its total action is thus the quadratic combination
$\frac{1}{2}(h,h)_\nabla-\gamma h_\varepsilon(z)$. Owing to (\ref{hf}) and to $\Var\, h_\varepsilon(z)=
(f_{\varepsilon}^z,f_{\varepsilon}^z)_\nabla$, the latter can be rewritten as
$\frac{1}{2}(h',h')_\nabla-\frac{\gamma^2}{2} \Var\, h_\varepsilon(z),$
with the substitution $h':=h-\gamma f_{\varepsilon}^z$. The probability weight involved in our random metric can therefore be written as
\begin{eqnarray}
\exp\left[-\frac{1}{2}(h',h')_\nabla\right]\mathbb E\, e^{\gamma h_\varepsilon(z)},
\label{quantumweight}
\end{eqnarray}
where the second factor  is the marginal distribution density (\ref{Eexphepsilon}) of $z$. The meaning of (\ref{quantumweight})
is that, after sampling $z$ from its marginal distribution, the law of $h$ \textit{weighted} by
$e^{\gamma h_\varepsilon(z)}$ is identical to that of the original GFF $h'$ \textit{plus} the deterministic
function
$\gamma f_{\varepsilon}^z$ (\ref{Newtonian}).

\textit{ KPZ proof.}
It is shown in \cite{2008arXiv0808.1560D} that when $\varepsilon$ is small,
the stochastic quantum measure (\ref{liouvillemeasure}) of the Euclidean ball $B_{\varepsilon}(z)$ is very well
approximated by
\begin{equation}
\label{mu}
\mu_\gamma(B_{\varepsilon}(z))\simeq \pi\varepsilon^{\gamma Q} e^{\gamma h_\varepsilon(z)},
\end{equation}
where $Q_\gamma:=2/\gamma+\gamma/2$.
  In the simplified perspective of this work,
we take (\ref{mu}) to be the {\em definition} of $\mu_\gamma(B_{\varepsilon}(z))$.  That is,
we view $\mu_\gamma$ as a function on balls of the form $B_{\varepsilon}(z)$, defined by
(\ref{mu}), rather than a fully defined measure on $D$.
Let us then call  \textit{quantum ball} $\tilde B^\delta(z)$
of area $\delta$ centered at $z$  the (largest) Euclidean ball $B_{\varepsilon}(z)$
whose radius $\varepsilon$ is chosen so that
\begin{equation}
\label{delta}
\mu_\gamma (\tilde B^\delta(z)) = \delta.
\end{equation}

One says that a (deterministic or random) fractal subset $X$ of
$D$ has \textit{Euclidean scaling exponent} $x$ (and {Euclidean dimension} $2-2x$) if,
for $z$ chosen uniformly in $D$ and independently
of $X$, the probability ${\mathbb P}\{B_\varepsilon(z) \cap X \not = \emptyset \} \asymp \varepsilon^{2x},$
in the sense
 that $\lim_{\varepsilon \to 0} {\log {\mathbb P}
}/{\log \varepsilon} =
2x.$ Similarly, we say that $X$ has \textit{quantum scaling exponent $\Delta$} if when
$X$ and $(z,h)$, sampled with weight (\ref{quantumweight}),  are chosen independently we have
\begin{equation}
\label{quantumP}
{\mathbb P}\{\tilde B^\delta(z) \cap X \not = \emptyset \} \asymp \delta^{\Delta}.
\end{equation}

In weight
(\ref{quantumweight}),
$h'=h-\gamma f_{\varepsilon}^z$ is a  standard GFF, thus its average has the
characteristic property (\ref{varh-h'}):
 $\mathcal B_t:=h'_{\varepsilon = e^{-t}}(z)$ is standard Brownian motion in time $t=-\log \varepsilon$.
  Eq. (\ref{Newtonian}) then gives
 $h_{\varepsilon}(z)=h'_{\varepsilon}(z)-\gamma \log \varepsilon=\mathcal B_t + \gamma t$ (up to a bounded constant), i.e.,
 $h_{\varepsilon}(z)$ in (\ref{mu}) sampled with (\ref{quantumweight}) has the same law as \textit{Brownian motion with drift}.

 Equality of (\ref{mu}) to (\ref{delta}) then relates stochastically the Euclidean radius $\varepsilon$ to the
 quantum area $\delta$. This radius is given in terms of the \textit{stopping time}
\begin{equation}
\label{stoppingtime} T_A=-\log \varepsilon_A:=\inf\{t: {-\mathcal B}_t + a_\gamma t=A\}
\end{equation}
with the definitions $A:=-(\log \delta)/\gamma >0$ and $a_\gamma:=Q_\gamma-\gamma={2}/{\gamma}-{\gamma}/{2}>0$
for $\gamma<2$.
A constant is absorbed in the choice of time origin
such that ${\mathcal B}_0=0.$

The probability
that the ball  $B_{\varepsilon_A}(z)$  intersects $X$ scales as $\varepsilon_A^{2x}=e^{-2xT_A}$.
Computing its expectation
$\mathbb E \exp{(-2xT_A)}$
with respect to the random time $T_A$ will give the quantum probability (\ref{quantumP}).
 Consider then for any $\beta$ the standard \textit{Brownian  exponential
martingale} ${\mathbb E}\left[\exp(- \beta {\mathcal B}_t - \beta^2 t/2)\right]=1$, valid for $0\leq t < \infty$.
 We can apply it at the stopping time $T_A$, when $T_A <\infty$ and where ${\mathcal B}_{T_A}=a_\gamma T_A -A$;
   we thus get  for $2x=\beta_\gamma a_\gamma +\beta_\gamma^2/2$
\begin{eqnarray} \nonumber
{\mathbb E}[
\exp(- 2x T_A) 1_{T_A<\infty}]&=&\exp(-\beta_\gamma A)=\delta^{\Delta_\gamma}\\
\label{fullexpmartingalekpz}
\beta_\gamma(x):=(a_\gamma^2+4x)^{1/2} -
a_\gamma,&&\Delta_\gamma:=\beta_\gamma/\gamma.
\end{eqnarray}
For $x=0$ one finds in particular
${\mathbb P}( T_A<\infty)={\mathbb E}[1_{T_A<\infty}]=1$, since  $\beta_\gamma(0)=0$ for $a_{\gamma< 2} > 0$,
so that the conditioning on $T_A<\infty$ can actually be omitted. We thus obtain the expected quantum scaling
behavior (\ref{quantumP}) with $\Delta=\Delta_{\gamma}$, which is the positive root to
 KPZ  (\ref{KPZrelation}), \textbf{ QED}.

The inverse Laplace transform $P_A(t)$ of (\ref{fullexpmartingalekpz}),
with respect to $2x$, is the
probability density of $T_A=-\log \varepsilon_A$ such that $P_A(t) dt:= {{{\mathbb P}}}
\left(T_A\in [t, t+dt]\right)$ \cite{2008arXiv0808.1560D}:
\begin{equation} \label{PA} P_A(t)=\frac{A}{\sqrt{2\pi t^3}} \exp\left[-\frac{1}{2t}\left({A}
-a_\gamma {t}\right)^2\right].
\end{equation}
From (\ref{PA}) one deduces that for $A$ large (i.e., $\delta$ and
$\varepsilon$ small), $\frac{A}{T_A}=\frac{\log \delta}{\gamma\log
\varepsilon_A}$ is concentrated in (\ref{fullexpmartingalekpz}) near
$a_\gamma + \gamma \Delta$. Reverse engineering to GFF $h$ via
(\ref{stoppingtime}), one finds that a point $z$ that is typical
with respect to the quantum measure is an $\alpha$\textit{-thick
point} of $h$ \cite{huperes}:
$\alpha:=\lim_{\varepsilon \to 0} \log h_\varepsilon(z)/\log
{\varepsilon}^{-1}=\gamma-\gamma \Delta,$
for a fractal of quantum scaling dimension $\Delta$.

\textit{Boundary KPZ.} Suppose that $D$ is a domain with a (piecewise) linear boundary $\partial D$
and  $h$ a GFF, now with \textit{free} boundary conditions. For $z\in \partial D$, $ h_\varepsilon(z)$
is the mean value of $h(z)$ on the \textit{semicircle} $\partial B_\varepsilon(z) \cap D$,  with
variance  scaling  like $-2\log
\varepsilon$. We define the boundary quantum
measure $d\mu^B_\gamma := \varepsilon^{\gamma^2/4} e^{\gamma
h_\varepsilon(z)/2}dz$, where now $dz$ is Lebesgue measure
on $\partial D$, with the conformal factor  needed
 for integrating a quantum length instead of an area, and a regulator such that the limit of $\mu^B_\gamma$ exists for
 $\varepsilon \to 0$ and $\gamma < 2$ \cite{2008arXiv0808.1560D}. For a fractal $X \subset \partial D$,
 we define boundary Euclidean ($\tilde x$) and quantum ($\tilde \Delta$) scaling exponents with this measure.
  We can repeat  the analysis above, with now  $ h_\varepsilon(z)$
  a standard Brownian motion $\mathcal B_{2t}$, with drift $\gamma t=-\gamma\log \varepsilon$, and
 prove the validity of KPZ relation (\ref{KPZrelation}) for the pair $(\tilde x,\tilde \Delta)$ \cite{2008arXiv0808.1560D},
 as anticipated in \cite{MR2112128}.

\textit{Liouville quantum duality.}
For $\gamma>2$, the Liouville measure (\ref{liouvillemeasure})
corresponds to the so-called ``other'' gravitational dressing of the
Liouville potential
\cite{1995PhRvD..51.1836K,1995NuPhB.434..264K,1996NuPhS..45..135K}.
The corresponding random surface is meant to be the scaling limit of
random simply connected surfaces with large amounts of area cut off
by small bottlenecks
\cite{1990MPLA....5.1041D,1992PhLB..286..239J,1992MPLA....7.3081K,1992PhLB..296..323K,1993NuPhB.394..383A,
1994NuPhB.426..203D,1994MPLA....9.1221A}.
This surface turns out to be a tree-like foam of
Liouville quantum bubbles of \textit{dual} parameter $\gamma' :=
4/\gamma$ (``baby-universes'') connected to each other at ``pinch
points'' and rooted at a ``principle bubble'' parameterized by $D$.
A precise description requires additional machinery and will appear
elsewhere. For now we relate $\gamma$ to $\gamma'$ only formally.

The definition of quantum balls in (\ref{mu}) and (\ref{delta}) makes
sense when $\gamma
> 2$.  Noting that $Q_{\gamma'}=Q_\gamma$, we have
$$\mu_{\gamma'}(B_\varepsilon(z)) = \pi \varepsilon^{\gamma' Q} e^{\gamma' h_\varepsilon(z)} =
\mu_{\gamma} (B_\varepsilon(z))^{\gamma'/\gamma} = \mu_{\gamma}
^{4/\gamma^2}$$ (up to an irrelevant power of $\pi$) --- i.e., a
$\gamma$-quantum ball of size $\delta$ has $\gamma'$-quantum size
${\delta}^{\,\prime}:=\delta^{4/\gamma^2}$.  (Intuitively, the ball
contains about a $\delta'$ fraction of the total $\gamma'$-quantum
area but only a $\delta < \delta'$ fraction of the $\gamma$-quantum
area because the latter also includes
points on
non-principle bubbles.) The number of $\gamma$-quantum size-$\delta$
balls needed to cover the principle bubble $D$ thus scales as
$(\delta')^{-1} = \delta^{-4/\gamma^2}$.

 From (\ref{KPZrelation}), the quantum scaling exponent $\Delta_\gamma$ in (\ref{fullexpmartingalekpz}), when
 {generalized} to $\gamma >2$,
 satisfies the \textit{duality relation} $\Delta_\gamma - 1 =
(4/\gamma^2)(\Delta_{\gamma'}-1)$ \cite{1995PhRvD..51.1836K,1995NuPhB.434..264K,1996NuPhS..45..135K,MR2112128}.
If $X\subset D$ has scaling
exponent $x$, then (\ref{quantumP}), established for $\gamma'<2$, essentially says (see
 \cite{2008arXiv0808.1560D})  that the expected number
$N_{\gamma'}(\delta', X)$ of $\gamma'$-quantum size-$\delta'$ balls (\textit{i.e.}, number
$N_{\gamma}(\delta, X)$ of $\gamma$-quantum size-$\delta$ balls)
required to cover $X$ scales as $(\delta')^{\Delta_{\gamma'}-1}=\delta^{\Delta_\gamma-1}$.

\textit{Brownian approach to duality.} When $\gamma>2$, the
$\varepsilon$\textit{-regularized} measures $M_{\varepsilon}(z)dz$
(\ref{liouvillemeasure}) converge to zero. 
If we choose the pair
$(z,h)$ from the weighted measure $M_\varepsilon(z) dh dz$ as in
(\ref{quantumweight}) and consider the Brownian description
(\ref{stoppingtime}), we find that $a_\gamma < 0$ for $\gamma>2$,
i.e., the drift term runs in a direction opposite to $A>0$, so that
$T_A = \infty$ for large $A$. The weighted measure is thus {\it
singular}; i.e., there is a quantum area of at least $\delta$
\textit{localized} at $z$ for small enough $\delta$. The Brownian
martingale result (\ref{fullexpmartingalekpz}) for $x=0$ gives the
probability, at a given $z$, for $T_A$ to be finite:
\begin{equation}\nonumber
\mathbb P(T_A<\infty)={\mathbb E}[1_{T_A<\infty}]=\delta^{\Delta_\gamma(0)}=\delta^{1-4/\gamma^2}=
\delta/\delta',
\end{equation}
where $\Delta_\gamma(0)=(|a_\gamma|-a_\gamma)/\gamma=1-4/\gamma^2$.
For general $x$,  (\ref{fullexpmartingalekpz}) scales as $\delta^{\Delta_\gamma}$.  We may define a
$\delta$\textit{-regularized}
 measure $M^\delta(z)dhdz$ as $M_{\varepsilon_A}(z)dhdz$ restricted to the event $T_A < \infty$.
Replacing $\gamma$ with $\gamma' = 4/\gamma$ and
$\delta$ with $\delta^{\,\prime}$ has the same effect as multiplying
$M^\delta(z)$ by ${\delta'}/{\delta}=\delta^{4/\gamma^2-1}$, so
$\frac{\delta'}{\delta}M^\delta(z)dz$ converges to $d\mu_{\gamma'}$.
This agrees with the \textit{conditional} expectation scaling as
\begin{eqnarray} \nonumber
\frac{{\mathbb E}[ \exp(- 2x T_A) 1_{T_A<\infty}]}{{\mathbb E}[1_{T_A<\infty}]}=
\delta^{\Delta_\gamma}\times\frac{\delta'}{\delta}
={\delta^{\,\prime}}^{\,\Delta_{\gamma'}}.
\label{normalizedfullexpmartingalekpz}
\end{eqnarray}

Using (\ref{fullexpmartingalekpz}) and $a_{\gamma'}=-a_\gamma$ one
obtains $\Delta_{\gamma} \Delta_{\gamma'} = x$, as anticipated in
\cite{MR2112128}. The typical GFF thickness
$\alpha=\gamma(1-\Delta_\gamma)=\gamma'(1-\Delta_{\gamma'})$ is
\textit{invariant} under duality
 and obeys the Seiberg bound $\alpha \leq Q$
\cite{1990PThPS.102..319S}; the \textit{string susceptibility
exponent}
  $\gamma_{{\textrm{str}}}=2-2Q/\gamma$ obeys the expected duality relation
  $(1-\gamma_{\textrm{str}})(1-\gamma'_{\textrm{str}})=1$
  \cite{1990MPLA....5.1041D,1992MPLA....7.3081K,1992PhLB..296..323K,1993NuPhB.394..383A,
  1994NuPhB.426..203D,1994MPLA....9.1221A,1995PhRvD..51.1836K,MR2112128,1995NuPhB.434..264K,1996NuPhS..45..135K}.
  Finally, for the $\textrm{SLE}_{\kappa}$ process
  $\gamma=\sqrt \kappa$  \cite{2008arXiv0808.1560D}, so that the Liouville $\gamma\gamma'=4$ and SLE $\kappa\kappa'=16$
  dualities coincide.

\begin{acknowledgments}
We  thank O. Angel, J.
Franchi, P. Jones, I. Klebanov, I. Kostov, G.-F.
Lawler, A. Okounkov and the late Oded Schramm for useful discussions.
\end{acknowledgments}
\bibliography{dkpzprl}
\bibliographystyle{unsrt}
\end{document}